# Comment on: Observation of a first order phase transition to metal hydrogen near 425 GPa.

Isaac F. Silvera[1] and Ranga Dias[2]
[1]Lyman Laboratory of Physics, Harvard University, Cambridge, MA 02138,
[2]Department of Physics and Astronomy, University of Rochester, Rochester, NY

Loubeyre, Occelli, and Dumas (**LOD**) [1] claim to have produced metallic hydrogen (**MH**) at a pressure of 425 GPa, without the necessary supporting evidence of an insulator to metal transition. The paper is much ado about nothing. Most of the results have been reported earlier. Zha, Liu, and Hemley [2] studied hydrogen at low temperature up to 360 GPa in 2012; they reported absorption studies up to 0.1eV. Eremets et al [3] studied dense hydrogen up to 480 GPa using standard bevel diamonds. They reported darkening of the sample and electrical conductivity in which they reported semi-metallic behavior around 440 GPa. In 2016 Dias, Noked, and Silvera [4] reported hydrogen was opaque at 420 GPa. In 2017 Dias and Silvera observed atomic metallic hydrogen at 495 GPa in the temperature range 5.5-83 K [5].

LOD did not acknowledge or discuss any earlier work relative to their observations. They report a high-pressure phase diagram that strongly disagrees with the existing literature. They fail to acknowledge that MH has already been made at a pressure of 495 GPa, and categorize this with claims that have been established as controversial [6,7], evidently to "muddy" the waters. We first clarify this situation. In 1989 Mao and Hemley claimed to observe MH at a pressure above 200 GPa from optical studies [6], but this was criticized as having no evidence of metallization [8]. They continued with optical studies and claims of MH [9] at even lower pressures; eventually it was conceded that there was no metallization in hydrogen or deuterium up to pressures of 216 GPa [10]. In 2011 Eremets and Troyan studied hydrogen to 300 GPa and claimed metallization [7]. Although they discovered two important new phases of molecular hydrogen, there was no evidence of MH [2,11] and eventually they conceded [12], showing that hydrogen was still transparent in the IR up to ~300 GPa. There were criticisms of the observation of MH by Dias and Silvera [13-16], but all of these objections have been answered [17]. Our observation has almost been corroborated by Eremets et al who achieved a pressure of 480 GPa [3] (see also Fig. 10, ref. [18]). Thus, the observation of MH at 495 GPa remains as an established achievement that has been defended against criticism. The paper by LOD claiming to have made MH has many problems and unsubstantiated claims that we now discuss.

**Claim of Metallic Hydrogen**

Let us first consider the claim by LOD of MH. In our Fig. 1 we reproduce their Fig. 1a images of their sample, along with images shown by Dias and Silvera [5] in their Fig. 2. Although materials that are not metallic can be shiny in the visible spectral region (for example Si or Ge), no metals with a significant plasma frequency are black in the eletromagnetic spectrum; free-electrons result in reflected light.

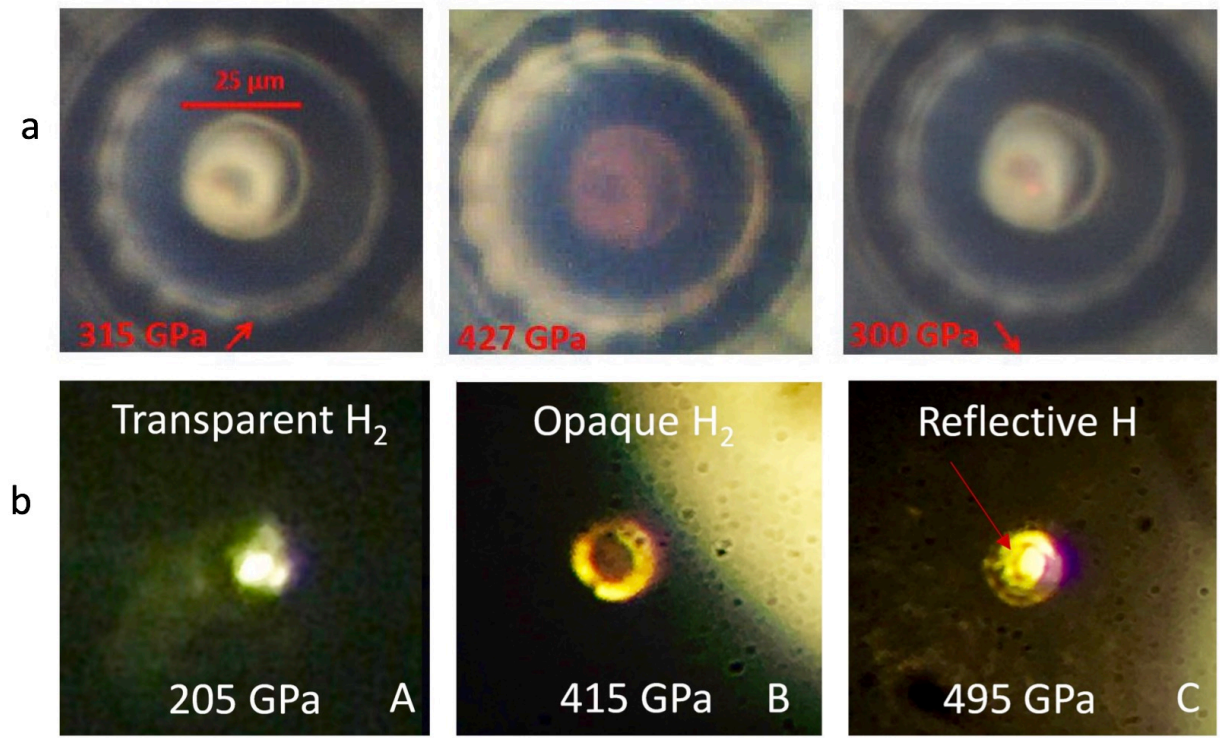

**Fig. 1.** a) Images of hydrogen from LOD in reflected light. All three photos show a small dark or black sample near the middle; in particular the middle image is claimed to be MH. We estimate that the linear dimensions of the sample are 5-6 microns. b) Images from Dias and Silvera. The left image is back lit and $H_2$ is transparent; the middle photo shows an opaque sample that is black in reflection; the bright ring around the sample is the Re gasket under the 30 micron diameter diamond culet. The right figure shows reflective atomic metallic hydrogen surrounded by the gasket; the diameter of MH sample is about 10 microns.

LOD's figure at 427 GPa is of modest resolution, but the sample appears to be black; the pink color surrounding the sample is reflected light, in which the high energy blue components of light have been absorbed by the stressed diamond.

How does one determine that a sample is a metal? The two conventional methods used in high-pressure research are

- Show that the electrical conductivity remains finite in the limit that temperature, T, approaches zero K. This require inserting electrical wires into the sample.
- Show that the reflectance increases with increasing wavelength, as predicted for a free-electron metal. This behavior persists in the low temperature limit and does not require electrical wires (non-intrusive).

None of these were shown by LOD who only studied their sample at a temperature of 80 K. In the observation of MH at 495 GPa [5] reflectance was measured to determine that the sample had transformed to atomic metallic hydrogen in the temperature range 5.5-83 K.

LOD claim electronic bandgap closure as evidence of metallization. In their Fig. 2b they interpret their data as closing of the bandgap. At 427 GPa they plot an abrupt reduction of the intensity of transmission of light to zero. Their data in Fig. 2d only shows that the transmission in the IR region under study goes to zero, not closure of a bandgap; their figure implies that the intensity changed discontinuously. This might be the behavior of a metal, but there are other reasons why the intensity can go to zero, for example, a transition to a phase that is absorbing in the IR, but not metallic; a black (absorbing) insulator or semiconductor! The discontinuity of the intensity could be the signature of a transition to a new phase that is not metallic. A phase with a reststrahlen band could inhibit propagation of light in their IR region. Thus, "Zero" transmission is not proof of an insulator-to-metal transition. In our Science paper (see the middle panel of our Fig. 1b, above) we observed zero transmission of visible and IR light at 420 GPa. The front illuminated sample was black in reflectance and not a metal. Thus, LOD cannot claim metallic behavior based on their data, unless they believe (see Fig. 1a above, middle) that black metals exist.

Using unpublished data from our earlier experiments, we have reanalyzed our IR spectra and performed the same calculation that LOD did in their paper (their Fig. 2). Our Fig. 2, below, shows zero integrated intensity at 420 GPa. This figure is almost identical to Fig. 2d of LOD, which is the basis for their claim of MH. We have reproduced this behavior many times in our experiments, but we did not claim the sample to be MH, simply because zero transmission is insufficient evidence. It is remarkable that LOD did not measure the reflectance of the sample with the powerful synchrotron source of radiation or the 660 nm laser used to measure the diamond Raman.

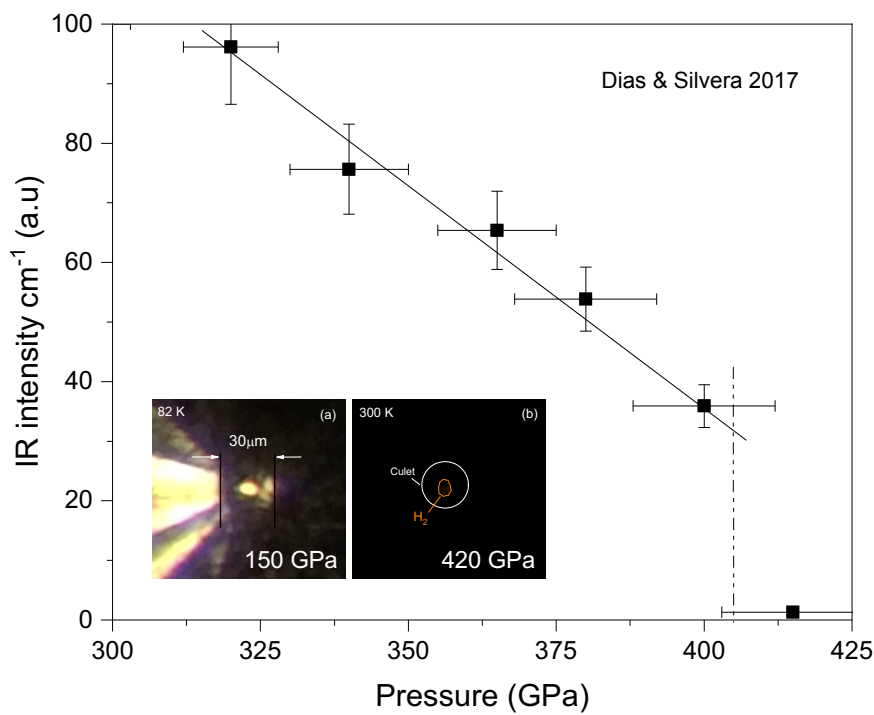

**Fig. 2.** Integrated transmitted intensity over the IR range (1800 cm$^{-1}$ to 3800 cm$^{-1}$) versus pressure, normalized to a 295 GPa pressure spectrum. The inset shows a photo of the darkening of the sample, around 420 GPa.

**Discussion**

A serious problem of the LOD paper is the disagreement with many other measurements of hydrogen as a function of pressure. In their paper they claim that hydrogen becomes black (in the visible) at 310 GPa (4040 cm$^{-1}$ on the hydrogen vibron pressure scale), whereas in an earlier paper [19] they found a value of 320 GPa (4025cm$^{-1}$). Only one other group has observed black hydrogen, and at a pressure of 300 GPa [20]. Already at the time of the original observation of black hydrogen this was quite controversial, as Narayanya et al [21] had observed hydrogen to be transparent at a higher pressure of 342 GPa. Silvera [22] commented on the conflict, suggesting that black hydrogen reported by Loubeyre et al [19] might be at a pressure 20% higher than they reported, due to their use of an older extrapolated calibration of the ruby scale. In our work, hydrogen was observed to darken and then become black above a pressure of ~360–370 GPa in three separate runs (T: 5.5–83 K); Zha et al [2] reported that hydrogen remains transparent at low and high temperatures; Eremets et al [3,23] reported darkening in two runs at P ~ 360 GPa for temperatures ~190 K . A problem with the LOD paper is that they seem to ignore much of the existing literature.  There are several observations at high pressuere in a DAC (Dias et al [24], 3 times; Zha, Liu, and Hemley [2], Eremets et al [3,12,23], Gregoryanz et al [25], Ruoff et al [21]). A few photos of samples are shown in Fig. 3. All had approximately the same thickness.

| RESEARCHER | IR VIBRON FREQUENCY OF HYDROGEN AT LOW TEMPERATURES (80-100K) | VISUAL OBSERVATION | CONCLUSION |
|---|---|---|---|
| Loubeyre et al (2002 and 2019) | 4040cm$^{-1}$ | Solid hydrogen was observed to turn **completely opaque** 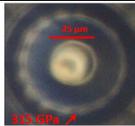 | Black hydrogen reported |
| Zha et al (2012 and 2013) | 4002cm$^{-1}$ | 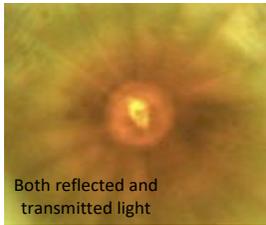 Both reflected and transmitted light | Sample still Transparent, **no black hydrogen** reported |
| Dias and Silvera (2016 and 2017) | 4005cm$^{-1}$ | 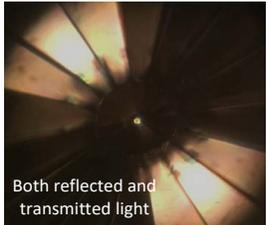 Both reflected and transmitted light | Sample still Transparent (Yellowish to Red in color) **no black hydrogen** reported |
| Eremets et al (2016 and 2019) | 337 GPa corresponding IR Frequency ~4035cm$^{-1}$ | 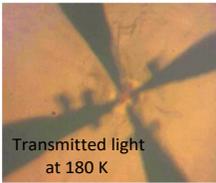 Transmitted light at 180 K | Sample still Transparent (Yellowish to Red in color) **no black hydrogen** reported |

Fig. 3. Some observations of hydrogen at high pressure in conflict with LOD. Pressures are determined by the peak frequency of the IR vibron. According to the calibration of the IR vibron peak versus pressure by Zha et al, peaks correspond to: 3988 (370 GPa), 4010 (359.3 GPa), 4025 (351.4 GPa). By contrast, LOD report a pressure of 320 or 310 GPa.

All other research groups observe transmission of hydrogen in the 310 to 350 GPa region; thus, not black at 300-320 GPa, in the visible. Two groups, the Harvard group and the Eremets group, have observed a phase transition at ~360 GPa and as the pressure is increased the transparent hydrogen darkens to black in reflectance and becomes non-transparent. Zha et al [2] also observe darkening at around 360 GPa. This darkening has been ascribed to the gap closing down and semiconducting behavior.

Another claim of LOD is that 400 GPa is the limit for conventional DACs, despite a multitude of counter-examples. One example from Eremets et al [3], using a conventional DAC is shown in Fig. 4; and of course Dias and Silvera achieved 495 GPa with a conventional DAC.

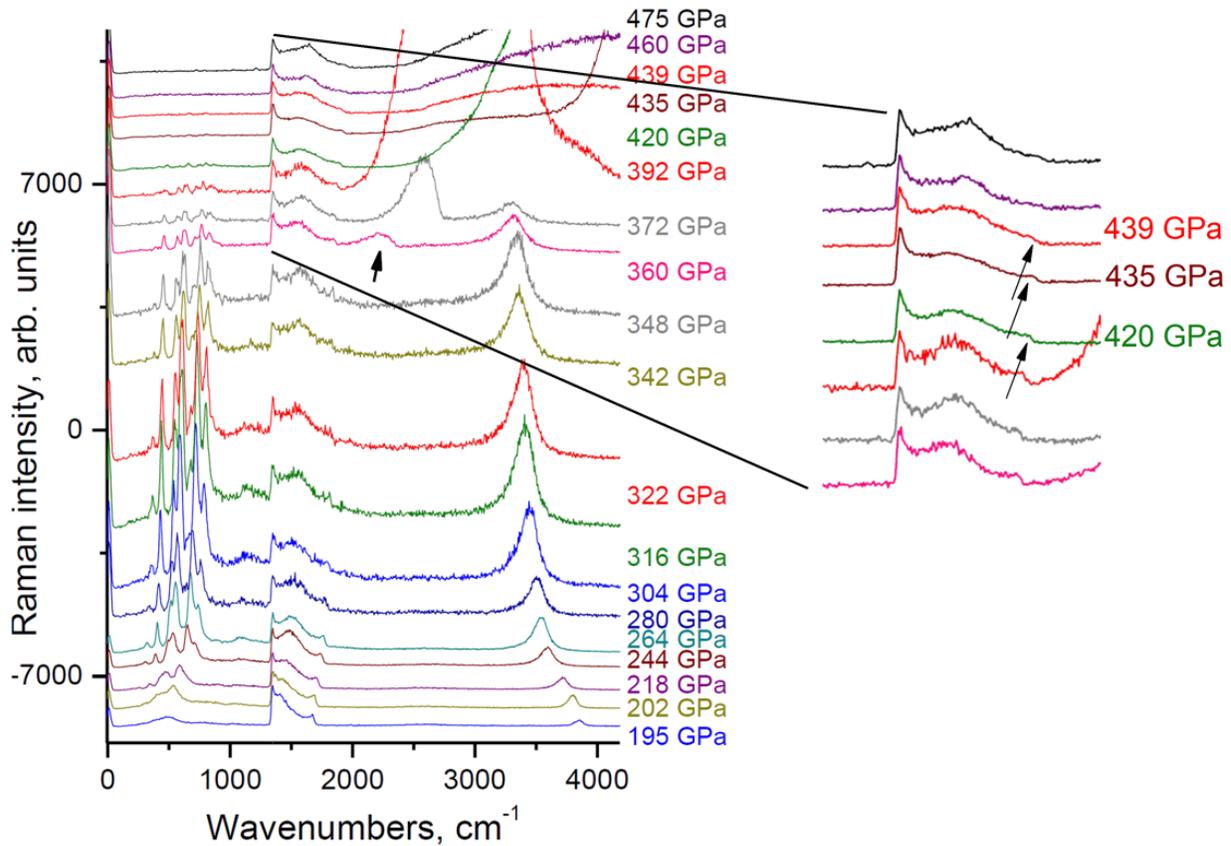

Fig. 4. Raman spectrum of diamond at various pressures, clearly showing pressures of more than 400 GPa using a standard diamond anvil cell.

**How can we understand why LOD's data differs from most of the literature?**

There are three possibilities that come to mind.

1. The first is that they have metallic impurities in their sample that affects their spectra. Hydrogen is known to be very reactive and diffusive at room temperature. At Harvard we use cryogenic loading to prevent diffusion and chemical activity of hydrogen; the activity rates increase exponentially with temperature. Thus, our hydrogen samples are solid and maintained at low temperatures to inhibit chemical activity. LOD load hydrogen under pressure in the liquid state at room temperature. Years ago, H. Shimizu et al [26] studied pressurized hydrogen by Brillouin scattering (BS). Later Shimizu and Kumazawa [27] continued the study and found that the metallic gasket confining the hydrogen dissolved impurities into the liquid hydrogen. They observed that at a fixed pressure the BS signal changed in time as the impurity concentration increased. <u>LOD actually show evidence of impurity modes in their Figure, Extended data 4</u>, so their samples are contaminated. A concentration of metallic impurities in hydrogen can impact the phase diagram of hydrogen, the vibron frequencies, and even induce metallization at a lower pressure than for the pure hydrogen crystal, as was studied theoretically by Carlsson and Ashcroft [28]. The fact that there is a scatter of pressures at which black hydrogen has been observed (300, 310, and 320 GPa) might be explained by varying concentrations of impurities. Thus, LOD's observation may be correct, but they were probably not studying pure hydrogen.
2. The second possibility is that the pressure scale of LOD, using the hydrogen vibron frequency, which can be affected by impurities, is incorrect. In Fig. 5, we plot the calibration data that we used along with that of Zha et al [2] and Hanfland et al [29], and that of LOD. The LOD calibration curve is distinctly different from the others, but all others agree with each other. The LOD calibration is evidently from unpublished data [13]. Extrapolating the LOD curve to lower pressures we find a large deviation of ~90 cm$^{-1}$ at a pressure where there have been many measurements of the vibron frequency. Extrapolating to higher pressures yields an interesting result. For a pressure of 320 GPa claimed by LOD, the other calibration curves give a pressure of ~360 GPa, in good agreement with other observations where hydrogen darkens. Clearly, there is a problem with the black hydrogen point. Apparently in their recent paper, they used a different pressure dependence of $H_2$ vibron.

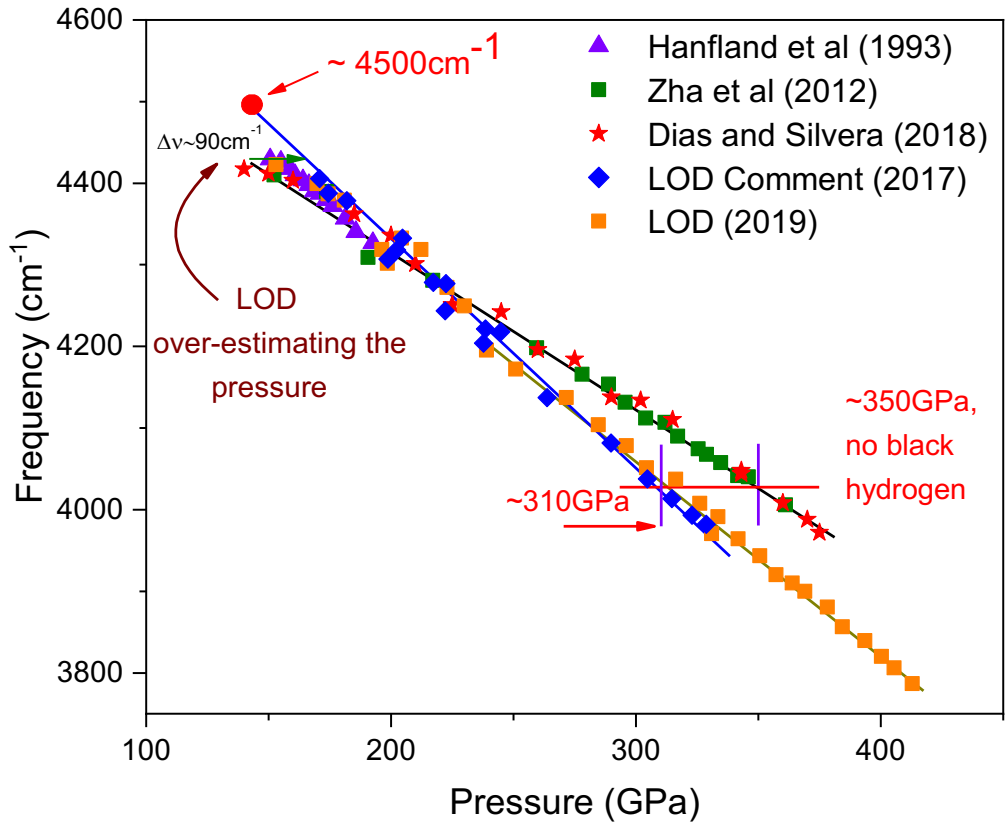

Fig. 5. Calibrations of the peak frequency of the IR vibron in hydrogen versus pressure showing the curve of Loubeyre et al (LOD) and other calibrations.

3. The third possibility is diffraction losses due to sample geometry. LOD neglect the effects of diffraction losses through their small sample hole in a metallic gasket. In general, when the wavelength of light is of the order of the hole size, or larger, the light will be cast out by Fraunhofer diffraction. Since they do not disclose the size of their sample, we estimate it based on the pictures provided. Using the culet size of 25 microns as reference, we estimate the longest dimension of their sample to be 5-6 micron. This is comparable to their longer wavelength IR region, and thus diffraction effects need to be taken into consideration in their analysis. To demonstrate how pronounced these effects can be, we use the well-known Airy disk equation:

$$I(\theta) = I_0 \left( \frac{2 J_1(\pi d/\lambda \sin(\theta))}{\pi d/\lambda \sin(\theta)} \right)^2$$

where $J_1$ is a Bessel function of the first kind, $\theta$ the angle of deflection, and $I_0$ the intensity of the zero'th order central bright peak. Looking at their geometry, treating the hydrogen sample as a pinhole, light passing through it will be cast out in the Airy diffraction pattern as defined above. LOD then capture a portion of this light in their Schwarzschild objective, to be carried through their optical system. This is shown in the Fig. 6, below:

To estimate the magnitude of diffraction effects, we can look at the amount of diffracted light that will be captured by their objective. LOD report a numerical aperture of 0.5, meaning the maximum angle of the light cone that can enter is 60 degrees ($\theta=30°$). The ratio of the integral of the Airy intensity function over the solid angle subtended by the objective to that over the entire half sphere gives an estimate of the fraction of transmitted light able to be received by their optical system (Fig. 7).

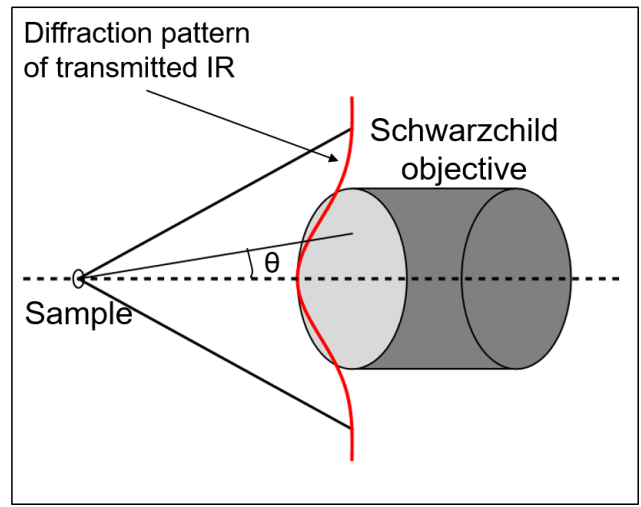

Fig. 6. LOD experimental geometry. Their sample can be considered as a pinhole, and the light passing through it will be cast out by diffraction, with intensity as a function of $\theta$ described by the Airy disk equation. We examine how much of the diffracted light will be captured by their objective.

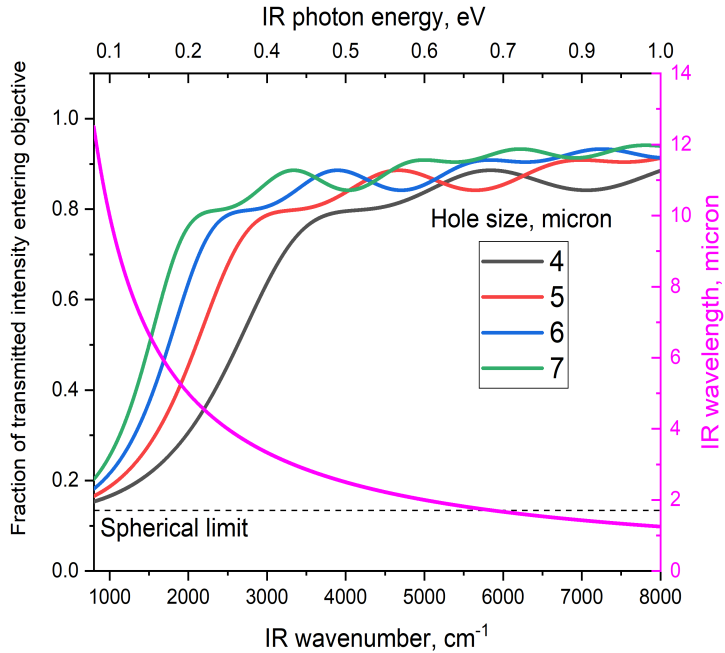

Fig. 7. Shown is the fraction of transmitted light caught by the Schwarzschild objective due to the transmitted light being diffracted outward by the sample hole. This fraction is plotted for several hole sizes, d. Included for reference is the conversion of the IR wavenumber to photon energy (top), and wavelength (pink curve). In the limit of λ>>d the emerging light will be a point source, and the fraction of transmitted light captured will be the ratio of the solid angle of the objective to the solid angle of the half sphere. This limit is shown by the dashed line.

As can be seen, when the IR wavenumber decreases and the wavelength of the IR becomes comparable to the sample size the fraction of transmitted intensity captured by the optical system rapidly drops, reaching the limit of spherical wavelets as $d/\lambda \ll 1$. This effect, particularly if the sample size changed as the pressure was increased, could dramatically change the interpretation of the presented IR absoprtion data.

A final comment concerns the phase diagram. At 425GPa, according to experiment, hydrogen is in the phase $H_2$-PRE, so perhaps what LOD observed is darkening of this phase coupled with small sample size. In their discussion of the possible phase diagram of hydrogen, they seem to cherry-pick the structures C2/c-24 and Cmca-12 for MH. Almost all advanced calculations of the structure of atomic MH predict the almost free-electron structure $I4_1/amd$ [30-34] which is ignored by LOD. See Fig. 8, the putative theoretical phase diagram.

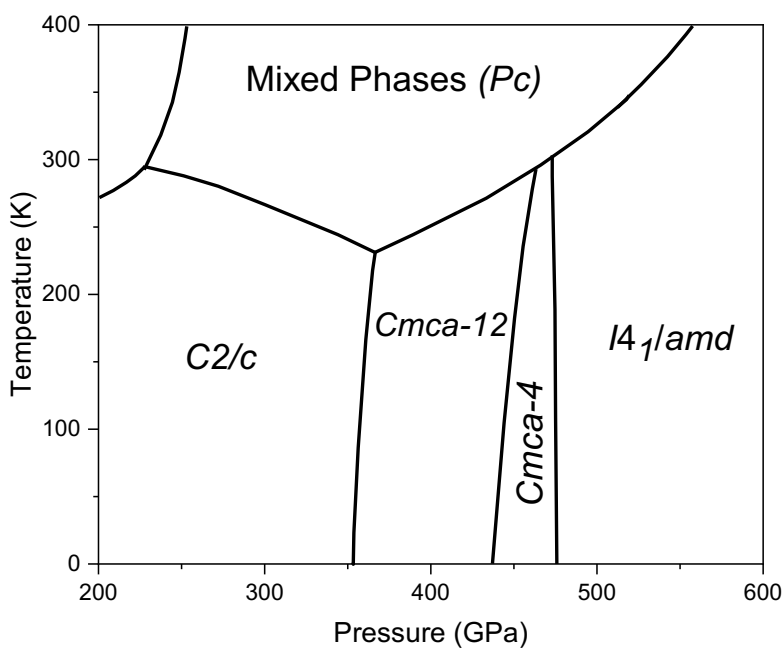

Fig. 8. Putative theoretical phase diagram of solid hydrogen

In conclusion, we believe that the paper by LOD is flawed, and their claim of metallization at 425 GPa may possibly be due to impurities in the sample and other possibilities, such as their deviant pressure scale, pointed out above. The observation of atomic MH in pure hydrogen at 495 GPa [5] does not suffer from these problems.